\documentclass[10pt,twocolumn,letterpaper]{article}
\usepackage[pagenumbers]{cvpr}
\usepackage{float}
\usepackage{array}
\usepackage{balance}

\definecolor{cvprblue}{rgb}{0.21,0.49,0.74}
\usepackage[pagebackref,breaklinks,colorlinks,allcolors=cvprblue]{hyperref}
\hypersetup{
    pdftitle={StructureClaw: Traceable LLM Agents and an Executable Benchmark for Structural Engineering Workflows},
    pdfauthor={Sizhong Qin, Yi Gu, Yao Jiang, Ao Cai, Changjian Zhou, Shaoxuan Shuai, Jiachang Wang, Tianhao Shen, Yueqiang Li, Xinhao Li, Li Zeng, Yueshi Chen, Dachen Gao, Genrong Xu, Wenjie Liao, Xinzheng Lu},
    pdfkeywords={large language models, structural engineering, agentic workflows, executable benchmark, traceable artifacts}
}
\def\confName{CVPR}
\def\confYear{2026}
\title{StructureClaw: Traceable LLM Agents and an Executable Benchmark for Structural Engineering Workflows}
\author{
Sizhong Qin$^\dagger$, Yi Gu$^\dagger$, Yao Jiang, Ao Cai, Changjian Zhou, Shaoxuan Shuai, \\ Jiachang Wang, Tianhao Shen, Yueqiang Li, Xinhao Li, Li Zeng, \\ Yueshi Chen, Dachen Gao, 
Genrong Xu, Wenjie Liao$^\ast$ and Xinzheng Lu$^\ast$
\\[3pt]
{\small $^\dagger$Equal contribution. $^\ast$Corresponding authors.}
}

\begin{document}
\maketitle

\begin{abstract}
Addressing a structural-engineering request requires more than a single answer; it requires a chain of interdependent artifacts: interpreted requirements, a computable model, validation records, solver outputs, applicable engineering checks, and a final report. Evaluations centered on question answering or script generation may therefore reward fluent outputs even when the underlying workflow is incomplete, inconsistent, or non-executable. We present StructureClaw, an artifact-centered workbench in which LLM agents operate through governed engineering skills, typed tools, shared artifact state, and local analysis backends, together with StructureClaw-Bench, an executable benchmark of 150 controlled scenarios spanning standard workflows, interactive robustness, and multimodal structural-model reconstruction. Its analyzable standard and multimodal cases require both strict one-to-one structural-model matching and numerical-response agreement with frozen reference responses from the selected analysis engine; interactive cases instead require positive clarification or recovery evidence together with safe non-execution when appropriate. A trial succeeds only when every fixture-required assertion passes. Across nine text-agent configurations, generic-only execution passed the model-artifact check in 87.0\% of retained outcomes but achieved only 22.0\% E2E Success, whereas automatic StructureClaw reached 82.9\%. Interactive and multimodal evaluations further identify semantic state consistency and executable model reconstruction as the dominant remaining bottlenecks. The code and benchmark are available at \url{https://github.com/structureclaw/structureclaw}.
\end{abstract}

\section{Introduction}
Large language models (LLMs) are increasingly applied to information retrieval, design assistance, and analysis in architecture, engineering, and construction (AEC) \cite{saka2024gptconstruction,kampelopoulos2025llmaec}. Structural-engineering agents, however, cannot be adequately evaluated as single-response systems. Engineering conclusions depend on a chain of mutually consistent artifacts, from interpreted requirements and structural models to analysis results, checks, and reports. A plausible answer or syntactically valid script may still hide missing or inconsistent intermediate states. Therefore, reliable engineering agents should expose and maintain the engineering evidence supporting their conclusions rather than only generate plausible endpoints. Throughout this paper, the evidence chain denotes the linked artifacts, tool executions, validation records, and terminal decisions that support an engineering result.

\begin{figure}[!t]
\centering
\includegraphics[width=\columnwidth]{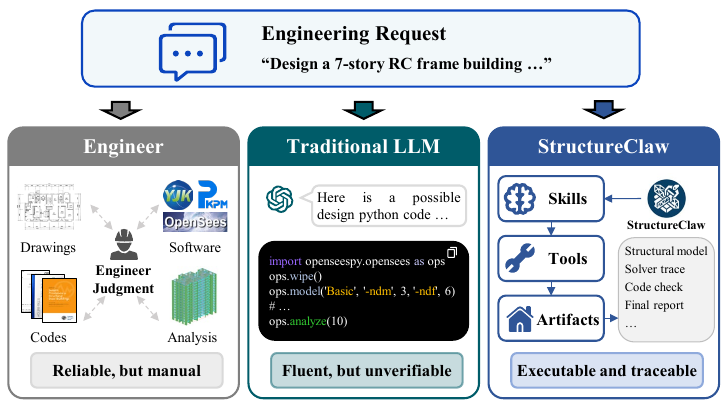}
\caption{From a structural-engineering request to reviewable evidence. Unlike a text-only response, StructureClaw preserves the skills, tool executions, and artifacts that support the reported result.}
\label{fig:teaser}
\end{figure}

We therefore argue that structural-engineering agents should be designed as workflow-level systems rather than isolated response generators. Their state should extend beyond conversational history to include persistent artifacts that can be inspected, validated, and reused throughout the workflow. This artifact-centered perspective enables engineering decisions to remain traceable and provides a foundation for reliable agent execution.

Recent studies have progressively extended structural-engineering automation from isolated design generation to analysis-integrated and agent-assisted workflows. Data-driven and generative approaches have increasingly incorporated structural analysis into the design process \cite{qin2025review,liao2021automated,lu2022physicsgan,zhao2023shearwallgnn,liao2024generative,gu2024diffusionshearwall,zhou2024structdiffusion}. LLM-based systems have begun to support interactive structural design, code-compliant design, and robust structural analysis \cite{qin2024llmshearwall,chen2025multiagentrc,liu2026llmstructuralanalysis}. Multi-agent orchestration and retrieval-augmented calculation have also been used to support task decomposition, tool coordination, and knowledge grounding \cite{liang2025masse,choi2026ramasc}. However, most existing studies focus on individual capabilities or partial workflows. A key remaining question is whether an engineering agent can maintain consistency across this evidence chain throughout an end-to-end workflow.

To address this challenge, we present StructureClaw, an artifact-centered workbench for structural-engineering agents that integrates governed skills and typed engineering tools. The defining design choice is to treat engineering artifacts, rather than conversational history alone, as the persistent state of the agent. Based on this principle, StructureClaw implements this design through a ReAct-style execution loop \cite{yao2023react}, a shared structural-model protocol, domain-specific skills, typed tools, backend providers, and persistent artifact records. These components allow intermediate models, validation results, solver outputs, and engineering checks to remain linked throughout execution.

\begin{figure*}[!t]
\centering
\includegraphics[width=0.96\textwidth]{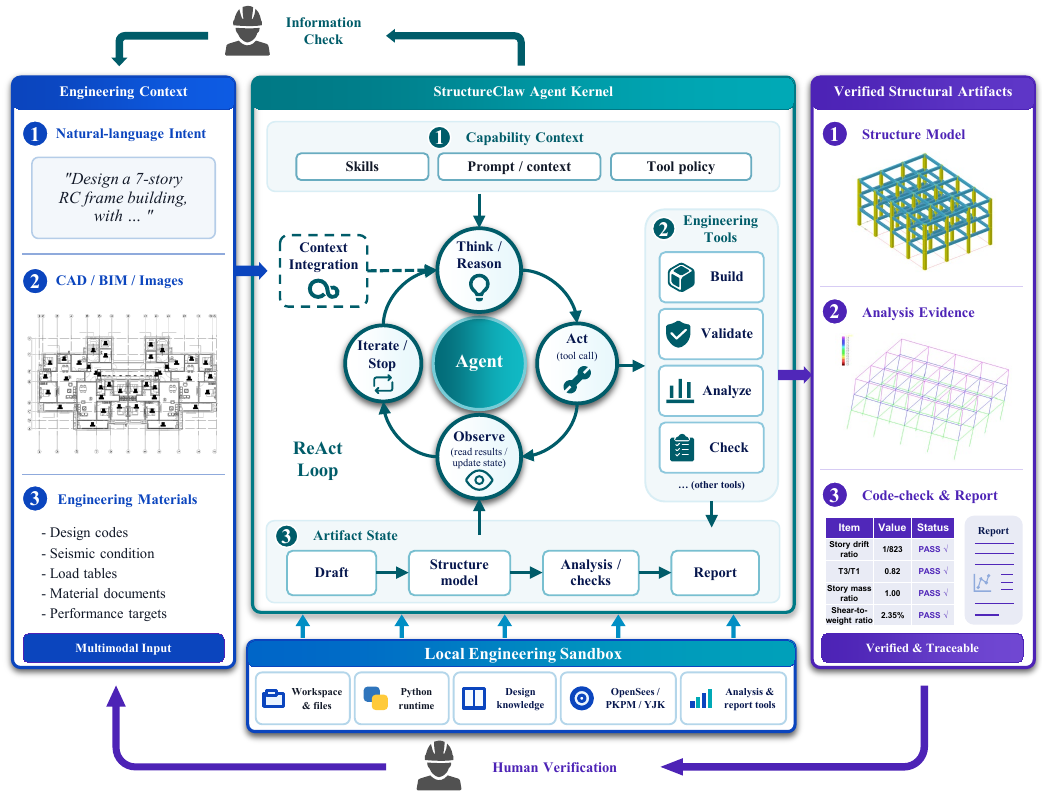}
\caption{Overall framework of StructureClaw. Multimodal engineering context enters a StructureClaw agent kernel, where capability context, engineering tools, and artifact state support a ReAct-style workflow. The local engineering sandbox executes model construction, validation, analysis, code checking, and report generation, producing structural artifacts and enabling validation-guided refinement.}
\label{fig:framework}
\end{figure*}

A traceable workbench alone does not establish reliability; evaluation must also execute and inspect the same evidence chain. OSWorld tests computer use \cite{xie2024osworld}, while $\tau$-bench targets policy-constrained tool interaction \cite{yao2025taubench}. AEC-Bench evaluates multimodal agentic tasks \cite{mankodiya2026aecbench}, whereas EngDesign uses simulation-validated engineering design \cite{guo2025engdesign}. These benchmarks, however, do not jointly evaluate strict structural-model fidelity, backend-aware numerical-response fidelity, safe non-execution, and trace-grounded reporting in one structural-engineering workflow. We therefore introduce StructureClaw-Bench, a controlled benchmark that operationalizes the evidence-chain formulation through scenario-specific, fixture-defined assertions. Its 150 scenarios cover standard workflow execution, interactive robustness, and multimodal structural-model reconstruction. The benchmark evaluates whether agents maintain a consistent evidence chain through executable assertions and repeated trials. Beyond measuring whether an agent can complete individual tasks, StructureClaw-Bench enables analysis of where reliability is lost along the engineering workflow. This perspective shifts evaluation from final-output quality toward the preservation of executable evidence across intermediate states.

In summary, this paper makes three contributions:
\begin{itemize}
    \item \textbf{Evidence-chain formulation of engineering agents.}
    We formulate structural-engineering agent reliability as the consistency of an executable evidence chain, requiring alignment among user intent, engineering artifacts, validation records, backend executions, terminal decisions, and final reports.

    \item \textbf{Artifact-centered engineering-agent workbench.}
    We develop StructureClaw, a workbench that realizes this perspective through persistent artifact state, governed skills, typed tools, a shared structural-model protocol, and backend-aware execution.

    \item \textbf{Executable benchmark and workflow-level insights.}
    We introduce StructureClaw-Bench, a 150-scenario executable benchmark, and conduct a large-scale diagnostic evaluation that localizes failure modes across workflow stages, including artifact construction, structural fidelity, interaction robustness, and multimodal reconstruction.
\end{itemize}
\section{Related Work}

\subsection{AEC LLM Systems}
AEC LLM systems have progressed from information access toward model authoring and executable engineering workflows \cite{kampelopoulos2025llmaec}. Recent systems explore BIM information access, model generation, and IFC-based authoring \cite{zheng2023bimsgpt,du2026text2bim,deng2025bimgent,nithyanantham2025mcp4ifc}; regulatory interpretation, code question answering, and model-based compliance \cite{madireddy2025codecompliance,iversen2026bimcompliance,joffe2025buildingcodes}; and structural-engineering tasks including code-compliant design, two- and three-dimensional analysis, foundation design, and multi-platform execution \cite{chen2025multiagentrc,liang2025automatedanalysis,geng2025frameagent,geng2026agentic3dframes,geng2026multiplatform,liu2026llmstructuralanalysis,youwai2026foundation}. Multi-agent frameworks further introduce task decomposition, verification, and solver routing \cite{liang2025masse,heo2026hybridstructural}. However, most existing systems evaluate specific tasks or final artifacts rather than maintaining consistency across an end-to-end engineering workflow. This motivates the need for an evaluation framework that considers artifact consistency, execution evidence, and workflow-level reliability.

\subsection{Tool-Augmented and Skill-Based Agents}

Tool-augmented LLMs provide a basis for executable reasoning and have been surveyed across autonomous-agent and tool-learning settings \cite{wang2024agentsurvey,xu2025toollearning}. ReAct interleaves reasoning and action, PAL delegates steps to programs, and Toolformer and Gorilla learn or select API calls \cite{yao2023react,gao2023pal,schick2023toolformer,patil2024gorilla}. ToolLLM and BFCL evaluate multi-step API and function use, while SkillAct and recent skill taxonomies study reusable agent behaviors \cite{qin2024toolllm,patil2025bfcl,liu2024skillact,jiang2026agenticskills}.

Domain-specific agent systems increasingly use structured skills or capabilities to constrain tool usage and execution behavior. StructureClaw follows this direction by defining governed skills with explicit interfaces, validation requirements, and artifact contracts.

\subsection{Executable Agent \& Engineering Benchmarks}

LLM evaluation increasingly uses process-aware environments. AgentBench and SWE-bench evaluate interaction and software repair \cite{liu2024agentbench,jimenez2024swebench}; OSWorld and $\tau$-bench test computer use and policy-constrained interaction \cite{xie2024osworld,yao2025taubench}. EngDesign covers simulation-validated design \cite{guo2025engdesign}. AECBench and DrafterBench evaluate AEC knowledge and civil-engineering task automation \cite{liang2026aecbench,li2025drafterbench}, while AEC-Bench and AECV-Bench target multimodal agentic tasks and engineering-drawing understanding \cite{mankodiya2026aecbench,kondratenko2026aecvbench}.

Existing benchmarks do not explicitly combine strict structural-model matching, backend-aware numerical validation, safe non-execution with positive clarification evidence, and trace-grounded reporting within a unified structural-engineering workflow. StructureClaw-Bench evaluates these requirements through fixture-defined assertions over observable intermediate and final outcomes. It uses typed assertions over intermediate and final artifacts so that end-to-end success and stage-specific failures can be measured in the same executable environment.

\section{StructureClaw System}
\label{sec:structureclaw-system}

StructureClaw organizes structural-engineering requests as transitions over persistent engineering artifacts rather than as a single final response. As shown in Figure~\ref{fig:framework}, natural-language requirements, project files, and engineering constraints enter an agent kernel combining capability context, a ReAct-style loop, typed tools, and artifact state. A local sandbox performs model construction, validation, analysis, supported checking, and reporting. Tool outputs update the shared state for subsequent actions and external review. The resulting artifacts are checked only within the system's supported scope and remain subject to professional verification.

\subsection{Skills, Tools, and Artifact State}

StructureClaw separates domain guidance, execution, backend binding, and state representation. A skill supplies planning guidance and artifact contracts and exposes one or more engineering capabilities. Typed tools perform bounded state transitions, providers bind those tools to concrete backends, and artifacts preserve engineering content and execution evidence across transitions. The agent selects applicable skills and tools from the request and current state, while declared contracts enforce expected inputs, outputs, and runtime preconditions.

The central engineering artifact is a shared structural-model protocol covering geometry and topology, materials and sections, supports, loads and combinations, analysis metadata, and units. Current providers include OpenSees/OpenSeesPy and configured PKPM and YJK backends \cite{mckenna2011opensees,zhu2018openseespy}. Provider availability is checked before execution, so an unavailable backend remains an explicit workflow condition rather than being replaced by an unverified result.

\subsection{Artifact-Grounded Validation and Recovery}

StructureClaw applies the ReAct reasoning--action pattern over artifact state \cite{yao2023react}. Each iteration may invoke a tool, request clarification, repair an artifact, or terminate; returned results update the evidence available for the next decision. Runtime validation checks artifact schema, references, completeness, finite values, supports, and load definitions. Blocking diagnostics trigger clarification, repair and revalidation, safe non-execution, or unsupported termination.

Together, these mechanisms impose a workflow invariant: downstream
actions must consume recorded artifacts rather than reconstructing
engineering state from conversational text. Validation records remain
linked to the model they check, analysis results retain their model and
provider dependencies, and reports are generated only from artifacts
available in the current state. Missing, invalid, or unsupported
conditions therefore remain explicit workflow outcomes instead of being
silently converted into downstream results. This stage separation makes
failures in routing, modeling, validation, execution, and reporting
independently observable, directly motivating the evidence-chain
assertions in StructureClaw-Bench. Detailed skill schemas, artifact
envelopes, provider contracts, validation semantics, and terminal states
are specified in the supplementary material.

\section{StructureClaw-Bench}
\label{sec:structureclaw-bench}
StructureClaw-Bench evaluates agents within executable engineering workflows rather than as standalone question-answering systems. Each scenario defines a required evidence chain, and success requires all corresponding assertions over routing, artifacts, execution, interaction, and reporting to pass. Beyond aggregate success rates, the benchmark preserves assertion-level outcomes to localize where an agent workflow becomes inconsistent. Plausible prose therefore cannot compensate for a missing or inconsistent engineering trace.

\subsection{Scenarios and Assertions}
\begin{figure}[!t]
    \centering
    \includegraphics[width=\columnwidth]{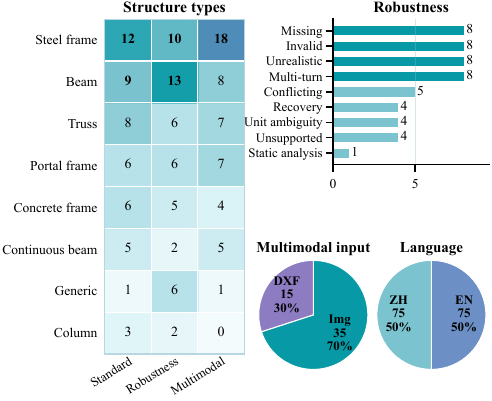}
    \caption{Composition of StructureClaw-Bench. The benchmark covers three workflow reliability dimensions: standard execution, interactive robustness, and multimodal structural-model reconstruction. The heatmap shows structure-type coverage, the bar chart summarizes interactive scenarios, and the pie charts show input-format and language distributions.}
    \label{fig:bench-distribution}
\end{figure}

The benchmark contains 150 controlled scenarios designed to evaluate three complementary reliability dimensions of structural-engineering agents, with 75 Chinese and 75 English prompts (Figure~\ref{fig:bench-distribution}). The standard-workflow family evaluates whether agents can transform well-specified requests into consistent analysis workflows for steel frames, concrete frames, beams, trusses, portal frames, continuous beams, columns, and generic structural-analysis tasks. The interactive-robustness family evaluates whether agents can identify missing information, invalid conditions, conflicting constraints, unsupported requests, and failed actions that require clarification, recovery, or safe non-execution. The multimodal structural-model reconstruction family evaluates whether agents can transform images or DXF inputs into executable structural artifacts by recovering geometry, topology, supports, loads, and member relationships.

Typed assertions operationalize the evidence chain into executable checks over intermediate and final artifacts. They evaluate workflow selection, artifact consistency, required analyses, safe behavior, and evidence-grounded reporting. Only fixture-required assertions are evaluated for each scenario, with reference fixtures defining reproducible operational targets.

Structural fidelity is measured through one-to-one mappings between generated and reference entities and their properties, while numerical fidelity compares mapped analysis responses against frozen backend results. These checks prevent plausible but inconsistent models or successful code execution without numerical consistency from being counted as success. Detailed validation procedures are provided in the supplementary material.

\subsection{Primary Metric and Evaluation Protocol}

Let $\mathcal{S}$ denote an evaluated scenario set, $\mathcal{A}_s$ the required assertions for scenario $s$, $\mathcal{R}=\{1,2,3\}$ the retained trials, and $z_{s,a,r}\in\{0,1\}$ the outcome of assertion $a$ in trial $r$. The primary metric is pooled single-attempt end-to-end (E2E) Success:

\begin{equation}
    \mathrm{E2E\ Success}(\mathcal{S}) =
    \frac{100\%}{|\mathcal{S}|\,|\mathcal{R}|}
    \sum_{r\in\mathcal{R}}\sum_{s\in\mathcal{S}}
    \prod_{a\in\mathcal{A}_s} z_{s,a,r},
    \label{eq:success-rate}
\end{equation}

where the product equals one only when every required assertion succeeds in a trial. This conjunctive formulation reflects the engineering setting in which a single missing or inconsistent artifact can invalidate an otherwise plausible result. Assertion-level outcomes are retained for diagnosis but provide no partial credit.

We additionally classify each scenario--configuration pair as stable success (3/3 trials), unstable (1/3 or 2/3), or stable failure (0/3). We do not report pass@3, because at-least-one success across repeated attempts would reward retrying rather than reflect the single-attempt deployment setting evaluated here.

Each scenario--configuration pair contributes three retained trial outcomes, allowing stable workflow reliability to be distinguished from stochastic success or failure.

The default automatic mode allows the agent to select applicable skills according to the request and artifact state. For standard-workflow scenarios, we additionally define a generic-only condition, in which the agent is restricted to the generic structural skill and the requested analysis skill while retaining the same execution environment. The resulting Auto--Generic comparison is a paired system-level analysis rather than a component-level ablation, because the two conditions jointly differ in specialized routing, structural priors, artifact expectations, and validation guidance.

\subsection{Diagnostic Views}

Beyond aggregate success measurement, StructureClaw-Bench supports failure localization through assertion-level diagnostics. Failed assertions are grouped into workflow stages, including routing, artifact construction, structural correctness, backend execution, numerical response, interaction safety, and reporting. Diagnostic rates use the complete fixture-applicable denominator, with unavailable required outcomes counted as failures under Equation~\ref{eq:success-rate}. We additionally stratify performance by scenario family, structure type, language, and modality. These analyses characterize failure patterns and support descriptive comparisons across the declared subsets.

\section{Experiments}
\label{sec:experiments}

We evaluate three benchmark families covering standard workflows, interactive robustness, and multimodal reconstruction. Each family contains 50 scenarios. Nine text-agent configurations are evaluated in the standard and interactive settings, and six agent--vision configurations in the multimodal setting. Standard scenarios are run in both automatic and generic-only modes; the other families use automatic mode. Three retained trials per condition yield 4,950 scored outcomes. A slash denotes agent / vision model when they differ. We additionally evaluate OpenClaw as an alternative agent-runtime baseline. It uses the same engineering tool implementations, analysis providers, benchmark scenarios, and evaluation protocol, while replacing the StructureClaw orchestration layer with a general-purpose agent loop.

\begin{figure*}[!t]
    \centering
    \includegraphics[width=\textwidth]{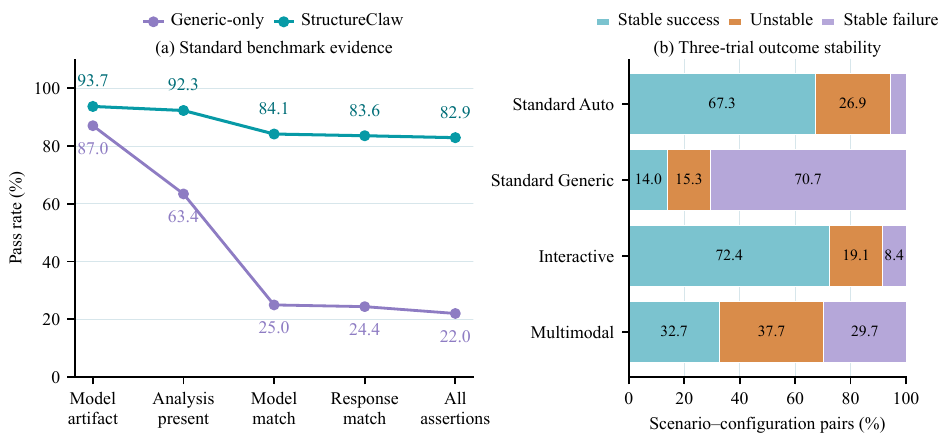}
    \caption{Evidence-chain and repeated-trial diagnostics. (a) Pooled standard-workflow rates across artifact availability, executable analysis evidence, structural fidelity, numerical-response fidelity, and the final all-required-assertions conjunction; every stage uses the complete 1,350-outcome denominator, with unavailable required evidence counted as failure. (b) Scenario--configuration pairs partitioned into stable success (3/3 trials), unstable (1/3 or 2/3), and stable failure (0/3).}
    \label{fig:evidence-stability}
\end{figure*}

Following Equation~\ref{eq:success-rate}, the primary metric is pooled single-attempt E2E Success; a trial succeeds only when every fixture-required assertion passes. The three-trial partition distinguishes stable success, unstable outcomes, and stable failure. Required but unavailable outcomes remain E2E failures, while assertion-level and resource measures are diagnostic. Model-level uncertainty is estimated with scenario-clustered bootstrap intervals. Because every configuration contributes the same number of attempts within a setting, pooled E2E rates equal the unweighted mean of configuration-level E2E rates. Figure~\ref{fig:evidence-stability} complements the model-level tables with evidence attrition and repeat stability; failure counts, input-format breakdowns, and resource measures are reported in the supplementary material.

\subsection{Standard Workflow}
\begin{table*}[!t]
    \caption{Three-trial standard-workflow results on 50 scenarios. E2E is pooled single-attempt success (\%); Stable is the 3/3 scenario rate (\%); Lift is the Auto--Generic improvement in percentage points. Structural fidelity denotes strict reference-model fidelity; Calls and Time are automatic-mode diagnostics. The Pooled row aggregates all StructureClaw configurations in the setting.}
    \centering
    \setlength{\tabcolsep}{2.5pt}
    {\small
    \begin{tabular*}{\textwidth}{@{\extracolsep{\fill}}lrrrrrrrr}
        \toprule
        Model & Auto E2E & Auto Stable & Generic E2E & Generic Stable & Lift & Structural fidelity & Calls & Time (s) \\
        \midrule
        GLM-5.2 & \textbf{97.3} & \textbf{92.0} & 36.7 & 30.0 & 60.7 & \textbf{98.0} & 6.52 & 196.0 \\
        GPT-5.5 & 91.3 & 86.0 & 36.0 & 30.0 & 55.3 & 91.3 & 7.03 & 139.4 \\
        Qwen3.7-Max & 89.3 & 78.0 & 23.3 & 12.0 & 66.0 & 90.0 & 6.75 & 187.6 \\
        Gemini 3.5 Flash & 88.7 & 70.0 & 4.0 & 0.0 & \textbf{84.7} & 94.7 & 6.11 & 45.4 \\
        Claude Opus 4.8 & 86.7 & 78.0 & \textbf{40.0} & \textbf{40.0} & 46.7 & 87.3 & 6.68 & 96.0 \\
        DeepSeek-V4-Flash & 83.3 & 66.0 & 22.7 & 8.0 & 60.7 & 84.0 & 6.69 & 104.6 \\
        DeepSeek-V4-Pro & 81.3 & 60.0 & 12.0 & 0.0 & 69.3 & 81.3 & 6.10 & 125.0 \\
        Kimi-K2.6 & 78.0 & 58.0 & 14.0 & 4.0 & 64.0 & 79.3 & 6.77 & 87.3 \\
        MiniMax-M3 & 50.0 & 18.0 & 9.3 & 2.0 & 40.7 & 51.3 & 7.20 & 56.5 \\
        \midrule
        \textbf{Pooled} & 82.9 & 67.3 & 22.0 & 14.0 & 60.9 & 84.1 & 6.65 & 115.3 \\
        \midrule
        OpenClaw + GLM-5.2 & 96.7 & 90.0 & 36.7 & 32.0 & 60.0 & 98.0 & 7.39 & 381.5 \\
        \bottomrule
    \end{tabular*}
    }
    \label{tab:standard-results}
\end{table*}

Automatic execution raises pooled E2E Success from 22.0\% to 82.9\%, a 60.9-point gain. All nine model-wise lifts are positive; every paired 95\% scenario-bootstrap interval is strictly positive, and the Holm-adjusted sign-flip $p$-value is 0.0009 for each comparison. GLM-5.2 has the highest observed automatic E2E Success at 97.3\%; the cross-model finding, however, is the positive lift for every configuration. Because automatic and generic-only modes jointly differ in routing, structural guidance, artifact expectations, and validation, this comparison supports the complete workflow configuration rather than a component-level causal effect.

The evidence view explains where the gain emerges. With unavailable required evidence counted as failure, generic-only execution passes the model-artifact assertion in 1,175/1,350 outcomes (87.0\%) but falls to 25.0\% at strict structural fidelity and 22.0\% E2E Success. Automatic execution retains 84.1\% structural fidelity and 82.9\% E2E Success, while stable success rises from 14.0\% to 67.3\%. Artifact presence alone therefore overstates workflow reliability. The OpenClaw comparison provides a controlled evaluation of orchestration efficiency. With the same engineering capabilities and evaluation protocol, OpenClaw achieves comparable E2E Success but requires more tool calls and longer execution time, suggesting that StructureClaw's domain-specific orchestration reduces execution overhead.

\subsection{Interactive Robustness}

\begin{table*}[!t]
    \caption{Three-trial interactive-robustness results on 50 scenarios. E2E is pooled single-attempt success (\%); Stable and Unstable denote the 3/3 and 1--2/3 scenario rates, while stable failure (0/3) is the unreported complement; the remaining rates are fixture-applicable diagnostics (\%). The Pooled row aggregates all StructureClaw configurations in the setting. }
    \centering
    \setlength{\tabcolsep}{2.5pt}
    {\small
    \begin{tabular*}{\textwidth}{@{\extracolsep{\fill}}lrrrrrrrr}
        \toprule
        Model & E2E & Stable & Unstable & Clarification & Safe model abstention & Safe analysis abstention & Calls & Time (s) \\
        \midrule
        GLM-5.2 & \textbf{90.0} & \textbf{82.0} & 14.0 & 92.9 & 97.1 & 98.0 & 4.73 & 105.4 \\
        Kimi-K2.6 & 88.7 & 80.0 & 14.0 & 96.0 & 99.0 & 99.0 & 4.20 & 79.7 \\
        Qwen3.7-Max & 85.3 & 78.0 & \textbf{12.0} & 98.0 & \textbf{100.0} & \textbf{100.0} & 4.56 & 95.2 \\
        Gemini 3.5 Flash & 84.7 & 74.0 & 20.0 & 96.0 & 97.1 & 97.1 & 4.87 & 22.5 \\
        Claude Opus 4.8 & 84.7 & 72.0 & 26.0 & \textbf{100.0} & \textbf{100.0} & \textbf{100.0} & 4.21 & 51.1 \\
        GPT-5.5 & 81.3 & 74.0 & \textbf{12.0} & 97.0 & 97.1 & 97.1 & 5.25 & 67.0 \\
        DeepSeek-V4-Pro & 80.7 & 70.0 & 22.0 & 93.9 & 97.1 & 98.0 & 4.88 & 73.7 \\
        DeepSeek-V4-Flash & 76.0 & 66.0 & 18.0 & 97.0 & 97.1 & 97.1 & 4.79 & 59.5 \\
        MiniMax-M3 & 72.0 & 56.0 & 34.0 & 91.9 & 95.1 & 95.1 & 5.59 & 33.3 \\
        \midrule
        \textbf{Pooled} & 82.6 & 72.4 & 19.1 & 95.8 & 97.7 & 97.9 & 4.79 & 65.3 \\
        \midrule
        OpenClaw + GLM-5.2 & 89.3 & 78.0 & 20.0 & 93.9 & 99.0 & 99.0 & 4.97 & 224.7 \\
        \bottomrule
    \end{tabular*}
    }
    \label{tab:interactive-results}
\end{table*}

\begin{table*}[!t]
    \caption{Three-trial multimodal-reconstruction results on 50 image/DXF scenarios. E2E, Stable, Image, and DXF are rates (\%); Type recognition and Structural fidelity use the complete fixture-applicable denominator, with unavailable required outcomes counted as failures. The Pooled row aggregates all StructureClaw configurations in the setting.}
    \centering
    \setlength{\tabcolsep}{2.5pt}
    {\small
    \begin{tabular*}{\textwidth}{@{\extracolsep{\fill}}lrrrrrrrr}
        \toprule
        Agent / vision & E2E & Stable & Image & DXF & Type recognition & Structural fidelity & Calls & Time (s) \\
        \midrule
        Gemini 3.5 Flash & \textbf{76.7} & \textbf{56.0} & \textbf{74.3} & 82.2 & 89.4 & \textbf{80.9} & 5.86 & 37.5 \\
        Claude Opus 4.8 & 68.7 & 40.0 & 61.0 & 86.7 & 90.1 & 68.8 & 5.62 & 92.0 \\
        GLM-5.2 / GLM-4.6V & 64.0 & 46.0 & 51.4 & \textbf{93.3} & 80.1 & 65.2 & 6.14 & 415.8 \\
        GPT-5.5 & 63.3 & 42.0 & 54.3 & 84.4 & \textbf{97.9} & 61.7 & 5.73 & 136.1 \\
        MiniMax-M3 & 33.3 & 12.0 & 32.4 & 35.6 & 85.8 & 31.2 & 5.68 & 49.4 \\
        Kimi-K2.6 & 13.3 & 0.0 & 6.7 & 28.9 & 35.5 & 13.5 & 4.77 & 45.2 \\
        \midrule
        \textbf{Pooled} & 53.2 & 32.7 & 46.7 & 68.5 & 79.8 & 53.5 & 5.72 & 129.3 \\
        \midrule
        OpenClaw (GLM-5.2 / GLM-4.6V) & 59.3 & 40.0 & 56.2 & 66.7 & 82.3 & 58.9 & 8.11 & 476.3 \\
        \bottomrule
    \end{tabular*}
    }
    \label{tab:multimodal-results}
\end{table*}

Interactive robustness reaches 82.6\% pooled E2E Success, with model rates from 72.0\% to 90.0\%. GLM-5.2 has the highest observed rate at 90.0\%, followed by Kimi-K2.6 at 88.7\%, illustrating that model ordering differs from the standard workflow. Stable success is 72.4\%, and each clarification or safe-abstention diagnostic exceeds 95\%. The gap between these local safety checks and E2E Success shows that clarification or abstention alone does not guarantee complete workflow consistency. For example, Claude Opus 4.8 achieves perfect clarification and abstention diagnostics but not the highest E2E Success, indicating that interactive reliability also requires maintaining artifact state consistency and completing downstream transitions.

Semantic evaluation is the most common first failed stage, followed by structural fidelity and interaction handling. These labels localize the first inconsistent handoff under the benchmark rules; they do not causally explain model reasoning.

\subsection{Multimodal Reconstruction}

Multimodal reconstruction reaches 53.2\% pooled E2E Success, with configuration-level rates from 13.3\% to 76.7\%. Gemini 3.5 Flash has the highest observed rate at 76.7\%, whereas Kimi-K2.6 falls to 13.3\% despite ranking near the top on interactive robustness. Model ordering is therefore task dependent. Because some rows combine different agent and vision models, these values compare complete configurations rather than isolated perception models.

Type recognition reaches 79.8\%, but strict structural fidelity reaches only 53.5\%, and stable success is 32.7\%. Structural fidelity is also the most common first failed stage. The dominant challenge is therefore converting perceived geometry into a complete executable model with correct topology, restraints, properties, and loads. Input-format differences remain descriptive because the image and DXF subsets contain different structures.

\section{Discussion}
\label{sec:discussion}

The results support workflow-level evaluation rather than artifact-level scoring. Generic-only execution often creates a model but loses downstream fidelity, whereas automatic execution improves E2E Success for every evaluated model. The OpenClaw comparison suggests that orchestration affects workflow efficiency beyond engineering capabilities, with comparable success but lower execution overhead under the same tools, skills, and protocol.

The model comparison further shows that reliability is task dependent. A configuration that performs strongly on well-specified or interactive text cases may perform poorly on multimodal reconstruction. Pooled results establish the workflow-level pattern, while model-level rows reveal configuration dependence; multimodal rows should not be read as isolated vision-model rankings.

Failure patterns identify different bottlenecks across task families. Standard workflows are limited mainly by structural fidelity, interactive tasks by semantic and state-consistency requirements, and multimodal tasks by executable reconstruction. High local rates for artifact presence, clarification, abstention, or type recognition therefore do not guarantee end-to-end success. Future systems should expose and validate intermediate engineering representations before execution and reporting.

Repeated trials also show that similar aggregate performance can hide case-level instability, especially in multimodal reconstruction. Reporting single-attempt E2E Success together with stable success, unstable outcomes, and stable failure is therefore more deployment-relevant than retry-oriented metrics.

The evaluation claims are scoped to three trials over a designed suite, a system-level mode comparison, strict fixture matching, a fixed semantic judge, and multimodal configurations that may couple different agent and vision models. Within this scope, the results establish workflow-level reliability patterns under the declared fixtures and runtime settings. Extending the evaluation to nonlinear analysis, broader code compliance, and project-scale deployment is a natural next step.

\section{Conclusion}

StructureClaw organizes structural-engineering agents around persistent, inspectable artifacts, while StructureClaw-Bench evaluates whether those artifacts form a consistent and executable evidence chain. Across 150 controlled scenarios, the benchmark shows that artifact presence alone overstates reliability: the automatic workflow achieves 82.9\% E2E Success versus 22.0\% for generic-only execution, local safety behavior does not guarantee complete interactive success, and multimodal perception remains substantially easier than executable reconstruction. The broader contribution is a framework for designing and diagnosing traceable engineering agents, not a claim of autonomous design competence or a replacement for professional judgment.
\clearpage
\begin{small}
\setlength{\bibsep}{0pt}
\linespread{0.99}\selectfont
\bibliographystyle{ieeenat_fullname}
\bibliography{references}
\end{small}
\clearpage
\appendix
\raggedbottom
\section{Implementation and Trace Semantics}
\label{app:implementation}

StructureClaw is implemented around an explicit evidence chain rather than a single conversational output. The chain records what the agent was asked to do, which governed capability it selected, which engineering artifacts were constructed, whether those artifacts passed the available checks, which backend was invoked, and what evidence was available when the report was written. This appendix specifies how skill manifests, guidance scopes, runtime contracts, artifact records, and terminal workflow states organize that chain, and briefly documents a post-hoc audit of the skill artifacts.

\subsection{Evidence Chain}

A request and its optional project files enter a sequence of workflow stages: requirement interpretation and routing, model construction, validation, analysis or code checking, and reporting. A stage consumes the current workflow state and records a typed state update, an artifact, or an explicit condition that prevents safe continuation. Tool-call records associate each executable operation with its inputs, returned object, provider binding, or error. Consequently, an evaluator need not infer model creation or solver execution from fluent prose alone.

Table~\ref{tab:app-evidence-chain} summarizes the evidence available at each stage. The table describes an audit interface, not a claim that every task must traverse every row. A request that lacks material information can terminate with clarification before model construction; an unsupported provider can terminate before execution; and a task that does not require analysis can finish without a solver artifact.

\begin{table}[tb]
\caption{Evidence-chain semantics. Each stage adds inspectable evidence or an explicit reason that the next stage should not run.}
\centering
\setlength{\tabcolsep}{4pt}
{\small
\begin{tabular}{>{\raggedright\arraybackslash}p{0.18\columnwidth}>{\raggedright\arraybackslash}p{0.35\columnwidth}>{\raggedright\arraybackslash}p{0.30\columnwidth}}
\toprule
Stage & Recorded evidence & Continuation condition \\
\midrule
Input and route & User request, project inputs, inferred structural family, and selected capability & The user intent and a compatible capability are identified, and required information is available \\
Modeling & Nodes, members, properties, supports, loads, and associated model metadata when produced & A model is complete enough for the requested operation \\
Validation & Deterministic diagnostics and any repaired model state & Prescribed workflow resolves blocking diagnostics before execution \\
Execution & Selected provider, analysis or checking invocation, returned artifacts, and errors & The provider is available and returns the required evidence \\
Reporting & Report artifact linked to the available workflow state & Claims are written from the artifacts actually present \\
\bottomrule
\end{tabular}
}
\label{tab:app-evidence-chain}
\end{table}

The shared structural model is the principal handoff between stages. Recognition without a model is therefore distinguishable from model construction, and model construction is distinguishable from successful execution. The same separation prevents a report artifact from serving as evidence that an analysis occurred: numerical claims require an analysis result or supported check record in the trace.

\subsection{Skill Schema and Artifact Graph}
\label{app:staged-skill-schema}

A skill schema describes how a capability is discovered, what scope-specific engineering guidance it contributes, the conditions under which it can be selected, and the artifacts it consumes or produces. The manifest contains the metadata needed for discovery and compatibility, including declared capabilities, selection requirements and conflicts, execution role, and artifact dependencies. Separate guidance documents supply intent-, draft-, analysis-, and design-level instructions.

\begin{figure}[!t]
\centering
\includegraphics[width=0.92\columnwidth,pagebox=cropbox]{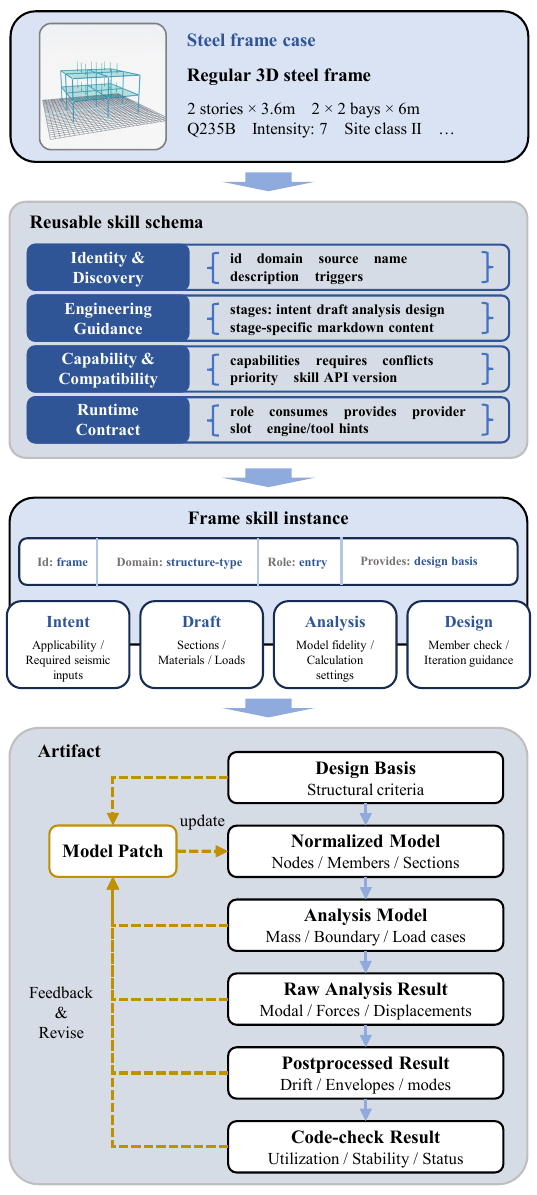}
\caption{Steel-frame example of the declarative skill and artifact protocol. The frame manifest defines reusable metadata and parallel guidance scopes, while typed tools and bound providers produce the request-specific artifact path. The lower panel shows one analysis-and-checking path through the artifact graph. Solid arrows denote required artifact dependencies. Dashed feedback indicates an optional model patch; after acceptance, base-revision validation, and conflict checks, the patch can produce an updated normalized model.}
\label{fig:app-skill-artifact-case}
\end{figure}

Figure~\ref{fig:app-skill-artifact-case} illustrates this schema for a regular three-dimensional steel-frame request. The corresponding manifest supplies frame-specific engineering guidance, serves as the entry capability for this structural family, and declares the design basis as its initial output. The intent, draft, analysis, and design labels identify parallel guidance scopes that can be activated as needed. Typed tools perform the requested operations, while separately selected providers perform analysis or checking. Their records establish which operations ran and which artifacts were returned.

\subsection{Artifact Envelope and Runtime Mapping}
\label{app:artifact-envelope}

During execution, StructureClaw stores each engineering object together with the metadata needed to trace how it was produced. The artifact contract distinguishes draft state, design basis, normalized model, validation record, analysis model, raw and postprocessed analysis results, code-check result, drawing artifact, and report artifact. Each record identifies its type and revision, producer and provider, run context, upstream dependencies, provenance, warnings, and type-specific engineering content. A normalized structural model, for example, contains units and project context; structural system and site conditions; nodes, elements, materials, sections, and boundary restraints; loads and combinations; analysis controls; metadata; and extensions. The contract defines more artifact types than the current runtime materializes, so a run records only those needed for the request.

The lower part of Figure~\ref{fig:app-skill-artifact-case} shows the dependencies among artifacts and model patches. A design basis records structural criteria such as material grade, seismic intensity, site category, and requested design scope. A model patch records proposed changes to an existing model. After its base revision has been verified and any conflicts resolved, an accepted patch is applied to produce a new normalized-model revision. Validation records contain deterministic diagnostics for that model. Analysis-model and raw-result artifacts contain mass, boundary conditions, load cases, eigenvalues, eigenvectors, forces, and displacements. Postprocessed results provide periods, modal participation, drift, and envelopes, while code-check results record utilization, stability, and status against the selected design basis and model. The report links its claims to the evidence produced by the workflow.

Each artifact can carry a dependency fingerprint derived from its upstream references and provider bindings. A code-check result can thus be traced to the design basis, model, and postprocessed results used to produce it, while a later patch creates a new revision without altering the earlier record. The dashed feedback in Figure~\ref{fig:app-skill-artifact-case} indicates that the protocol allows such a patch; the benchmark does not execute this loop automatically. In the current runtime, draft, model, validation, analysis, code-check, and report payloads are stored in separate state channels, and tool calls emit compact summaries.

\subsection{Provider Binding and Execution}

Execution remains tool-controlled. The agent can clarify, construct, validate, repair, execute, or stop. Computation begins only when a valid model is passed to an available provider; a missing provider or invalid model blocks execution.

Provider identity is part of the execution evidence. Analysis-provider contracts consume an analysis model and provide raw results; code-check-provider contracts consume the design basis, normalized model, and postprocessed results required by the selected check. Availability is checked before a provider-dependent action, and the selected analysis engine or design-code capability can be compared with scenario expectations. A completed analysis includes a recorded solver invocation and the artifact returned by the selected backend. Correctness still depends on the adapter, solver, units, and input model.

The automatic and generic-only settings alter the capability guidance available to the agent. The former exposes the structured selection path and its specialized artifact expectations. The latter retains the generic structural capability and the requested analysis capability but removes the automatic specialization path. The experiment therefore compares two complete workflow configurations; Appendix~\ref{app:protocol} specifies the execution matrix.

\subsection{Validation, Repair, and Terminal States}

The prescribed workflow places validation between artifact construction and provider execution. When a deterministic check exposes a correctable defect, the agent may repair the current artifact and validate it again within the same trial. This within-trial recovery is distinct from rerunning a completed benchmark case: each trial contributes one retained outcome, and separately logged infrastructure reruns do not create additional scored attempts. The workflow can therefore validate and repair an artifact before reaching a terminal state without changing the evaluation unit.

The trace distinguishes four terminal decisions. Completion means that the artifacts required by the request are available. Clarification means that material information remains missing or ambiguous. Safe non-execution means that analysis should not proceed because an input is invalid or inconsistent. Unsupported termination means that the requested provider or operation is unavailable in the evaluated environment. These states are judged from positive and negative evidence together. In particular, the mere absence of an analysis artifact is not a successful safety response: a negative case must also satisfy its scenario-specific clarification or semantic requirement and show that neither an analysis result nor an analysis-tool invocation was recorded. Appendix~\ref{app:benchmark} gives the corresponding assertion rules.

\subsection{Scope of Traceability}

The trace shows whether an action occurred, whether the expected artifact was present, and whether stage-local checks accepted it. These records help localize broken handoffs, such as correct routing followed by a missing model or a model followed by an absent analysis result. They provide stronger evidence than the final text alone because they refer to the executed state.

They are not by themselves proof of engineering correctness. The trace does not expose unrecorded reasoning, certify that a validator is complete, or guarantee that a reference fixture covers every engineering-equivalent solution. StructureClaw-Bench therefore augments trace evidence with strict reference-model matching and frozen numerical responses from the selected solver, while retaining explicit limits on alternative idealizations, solver validation, and professional scope. Construction suitability, exhaustive code compliance, and equivalence to professional review remain outside the evaluated claim.

\subsection{Skill Audit and Refinement}
\label{app:skill-audit}

We also conducted a post-hoc audit of the frozen skill library to identify opportunities for refinement. The audit covered 43 skills in 11 domains and treated each skill directory as a whole, including its manifest, stage guidance, declared resources, runtime, and tests where present. We prepared two prompts for each skill. The first described an ordinary supported request, while the second introduced missing information, ambiguity, or a capability boundary. This produced 86 prompt specifications.

Each skill was assessed on manifest quality, workflow clarity, failure handling, confirmation points, specificity, resource use, organization, prompt behavior, and prohibited actions. The resulting score was used to locate the most evident weakness and guide one focused revision. It did not determine whether that revision was accepted. Three independent judges instead compared the original and revised artifacts in the same evaluation context, and the majority decision determined whether the change was retained.

Failure handling was the largest initial weakness for 15 skills, followed by boundary-case behavior for 14, confirmation points for 10, and workflow organization for four. Most revisions clarified when execution should stop, what information should be requested, how a failed step should be reported, and which defaults or conclusions were not supported. The OpenSees static-analysis guidance, for example, added checks for rigid-body motion and mechanisms before solver execution. The GB50010 guidance separated the inputs required by each check from placeholder values that cannot support a compliance conclusion. In total, the audit revised 47 guidance files and one runtime. All 43 proposed changes received unanimous paired preferences and were retained.

\section{Benchmark Construction and Assertions}
\label{app:benchmark}

StructureClaw-Bench is a controlled fixture suite for evaluating workflow outcomes. Its unit of evaluation is a scenario, not an isolated question or a field project. Each scenario specifies an input, evaluation metadata, a reference fixture when artifact comparison is required, and the evidence that one single-attempt trial must produce or withhold.

\subsection{Scenario Specification and Coverage}

Conceptually, scenario $s$ is represented as
\begin{equation}
s = \left(x_s, m_s, f_s, \mathcal{A}^{\mathrm{req}}_s\right),
\label{eq:app-scenario-specification}
\end{equation}
where $x_s$ contains the prompt and any project file, $m_s$ records family, language, structural category, and input modality, $f_s$ is an optional reference fixture, and $\mathcal{A}^{\mathrm{req}}_s$ contains the typed assertions listed by the fixture. Equation~\ref{eq:app-scenario-specification} is a conceptual schema rather than an additional benchmark score. Every listed assertion contributes to the binary scenario outcome. The diagnostic views are marginal aggregations of these same assertion results and scenario metadata; they provide neither partial credit nor an alternative success criterion.

The evaluation suite contains 150 scenarios, allocated as shown in Table~\ref{tab:app-benchmark-coverage}. Each family is balanced between 25 English and 25 Chinese prompts, yielding 75 prompts in each language overall. The multimodal family contains 35 image cases and 15 DXF cases. This stratification provides controlled coverage of the three intended workflow regimes. Family, language, structure, category, and modality subsets support within-suite diagnostic analysis, while broader population-level transfer is considered through the validity boundaries in Appendix~\ref{app:diagnostics}.

\begin{table}[tb]
\caption{Scenario coverage. Language is balanced within every family; image and DXF counts refer to the multimodal family.}
\centering
\setlength{\tabcolsep}{4pt}
{\small
\begin{tabular}{>{\raggedright\arraybackslash}p{0.30\columnwidth}>{\centering\arraybackslash}p{0.11\columnwidth}>{\centering\arraybackslash}p{0.14\columnwidth}>{\raggedright\arraybackslash}p{0.29\columnwidth}}
\toprule
Family & Cases & EN/ZH & Input \\
\midrule
Standard workflow & 50 & 25/25 & Text \\
Interactive robustness & 50 & 25/25 & Text \\
Multimodal reconstruction & 50 & 25/25 & 35 image, 15 DXF \\
\midrule
Total & 150 & 75/75 & 100 text, 50 multimodal \\
\bottomrule
\end{tabular}
}
\label{tab:app-benchmark-coverage}
\end{table}

Standard-workflow scenarios provide sufficiently specified text requests and require the designated modeling, execution, and reporting path. Interactive scenarios introduce missing, conflicting, invalid, unrealistic, ambiguous, multi-turn, recovery, unsupported, and static-analysis conditions and test whether the agent clarifies, recovers, or withholds execution as specified. Multimodal scenarios add a perception-and-reconstruction stage before the same artifact checks. Keeping the families separate prevents performance on well-specified text requests from concealing unsafe execution or inaccurate reconstruction.

\subsection{Reference Fixtures}

The term reference fixture denotes the benchmark artifact against which a scenario is evaluated. It is a reproducible engineering target, not an assertion that one discretization is the unique correct design. The 59 model fixtures record a complete coordinate and unit contract, nodes, elements, materials, sections, supports, loads, load combinations, story information when applicable, and the analysis properties needed to reproduce the corresponding solver response. Scenario metadata can additionally specify accepted structural-type aliases, alternative capability identifiers, a required engine, or comparison tolerances.

The fixtures and numerical references are generated rather than manually transcribed. Every OpenSees reference model is solved in at least two fresh Python processes. The second path reorders model entities and loads, changes tag assignment, and uses a different equation system, numberer, and solution algorithm while preserving the engineering model. A truth set is written only when all results are finite, complete response trees agree within $10^{-10}$ absolute or relative tolerance, and global force and moment equilibrium pass. Canonical simply supported and cantilever beams are also checked against closed-form Euler--Bernoulli responses, and all 53 canonical two-dimensional fixtures are cross-validated using an independent direct-stiffness implementation. Six regular three-dimensional fixtures additionally retain pairwise OpenSees, PKPM, and YJK validation records. These cross-engine records diagnose agreement; engine-specific benchmark assertions continue to use frozen responses from the engine requested by the scenario.

Strict fixture matching detects both missing and extra entities, although valid engineering alternatives may use different discretizations, grouping conventions, idealizations, or design assumptions. The benchmark therefore measures agreement with its declared fixtures and operational rules rather than the uniqueness of the underlying design. Leakage guards additionally verify that production agent code contains no scenario identifiers or distinctive prompt fragments and that evaluators contain no per-scenario branches. Externally authored challenge cases would further test transfer across fixture conventions.

\subsection{Assertion Roles and Operational Rules}

For an evaluated configuration $c$, scenario $s$, trial $r$, and assertion type $a$, the recorded outcome is assigned one of four states:
\begin{equation}
o_{c,s,a,r}\in
\{\mathrm{pass},\mathrm{fail},\mathrm{inapplicable},\mathrm{unavailable}\}.
\label{eq:app-assertion-states}
\end{equation}
The vocabulary in Equation~\ref{eq:app-assertion-states} separates evaluation meaning from data availability. A pass or fail is a recorded result for an assertion evaluated on that run. Inapplicable means that the fixture does not request that assertion type; it is not a failure. Unavailable means that an expected observation could not be recovered because the run, trace, or result parsing was incomplete. An unavailable required outcome makes both the scenario and the corresponding assertion-level diagnostic unsuccessful, as formalized in Appendix~\ref{app:protocol}. This policy avoids conflating semantic inapplicability with missing evidence while preserving the complete applicable denominator.

Table~\ref{tab:app-assertion-rules} records the principal operational rules. Most rules are deterministic over the recorded agent state. The natural-language condition is the exception: a separate judge evaluates the scenario-specific semantic description with zero sampling temperature, a 4,096-token response limit, a 120-second timeout, and at most three protocol attempts. Exact reruns also depend on provider-side model behavior and the fixed evaluator configuration, as stated in Appendix~\ref{app:protocol}. Judge-dependent outcomes are reported separately from deterministic engineering assertions. Because the judge has not been calibrated against independent professional raters, these outcomes measure consistency with the fixed benchmark rubric rather than professional correctness.

\begin{table}[tb]
\caption{Principal assertion rules. Thresholds are operational definitions, not professional certification.}
\centering
\setlength{\tabcolsep}{4pt}
{\small
\begin{tabular}{>{\raggedright\arraybackslash}p{0.27\columnwidth}>{\raggedright\arraybackslash}p{0.59\columnwidth}}
\toprule
Assertion role & Recorded pass condition \\
\midrule
Structure recognition & The recorded structural type belongs to the fixture's accepted set. \\
Capability selection & The selected capability belongs to the fixture's allowed identifiers. \\
Model existence & A model artifact meets the scenario's minimum node and element counts; this stage tests artifact presence only. \\
Strict model matching & One-to-one node, element, and load matching passes together with coordinate, topology, restraint, assigned-property, modifier, story, and load-combination checks. \\
Numerical matching & The strict model match passes and mapped solver responses agree with frozen truth from the selected engine within the fixture tolerance. \\
Analysis and backend & A successful result contains displacement, reaction, or force data; backend matching additionally requires the recorded engine and analysis capability requested by the fixture. \\
Safe non-execution & No analysis result and no analysis-tool invocation are recorded; when required, no computable model is present and clarification or the specified semantic response also passes. \\
Report and semantics & Report presence requires more than 100 markdown characters; numerical provenance requires the structured report to embed the current analysis artifact; other scenario-specific language is evaluated by the fixed judge described above. \\
\bottomrule
\end{tabular}
}
\label{tab:app-assertion-rules}
\end{table}

\subsection{Strict Structural-Model Matching}

The formal experiments use a single strict matcher configuration. Strict matching first verifies the coordinate convention, model dimension, schema version, unit system, finite values, unique identifiers, cross-references, and load-combination references. It then finds a one-to-one coordinate-based node mapping and a one-to-one element mapping that preserves member type and mapped endpoint connectivity. Node, element, and load precision and recall must all equal one, so missing or additional entities fail.

Mapped nodes must have identical six-degree-of-freedom restraint vectors. Mapped elements are checked for assigned material and section properties, parametric section geometry, releases, offsets, rotation angles, grades, and declared reference vectors. Loads are matched by load-case type, mapped node or element target, load type, reference frame, direction, location, and magnitude; load combinations are compared by their resulting mapped load vectors. Story geometry, membership, and declared floor loads are also checked when present. Formal scenarios use a 0.05~m coordinate tolerance and 5\% relative tolerances for declared properties and load magnitudes. Entity identifiers may differ because the mappings preserve engineering relationships rather than string identity.

\subsection{Numerical and Report Fidelity}

All 50 standard scenarios and 47 multimodal scenarios evaluate numerical fidelity. The same 47 multimodal scenarios also evaluate structural-type recognition and strict structural fidelity, corresponding to the Type recognition and Structural fidelity diagnostics in the main results table. The remaining three multimodal scenarios do not request these reconstruction assertions and are semantically inapplicable to those diagnostics. Consequently, each multimodal configuration has $47\times3=141$ fixture-applicable retained outcomes for all three assertions; an unavailable required outcome remains in this denominator as a failure. Numerical matching is conditional on the strict model match, which supplies the node and element mappings required to compare responses without assuming identical entity identifiers. OpenSees assertions compare every available signed displacement, reaction, and member-response value recorded by the fixture. Reversed member orientation is normalized before comparing signed end forces. The scenario tolerance is 1\% relative with an absolute tolerance of $10^{-8}$ for near-zero values.

Eighteen standard scenarios explicitly request an analysis backend: six use OpenSees, six PKPM, and six YJK. PKPM and YJK assertions compare runtime results with frozen same-engine ground truth at 1\% tolerance. PKPM covers every available non-base-node displacement and member-force component; YJK additionally covers every node reaction. Exact entity coverage is required. OpenSees cross-engine results remain an independent validation reference and are not substituted for a commercial engine's own response.

Four scenarios require a report. Report-grounding results are therefore treated as a targeted diagnostic rather than benchmark-wide evidence of reporting reliability. For these scenarios, numerical consistency requires the structured analysis object embedded in the report to agree with the current analysis artifact at near-exact tolerance and to contain at least one numerical value. Report length establishes artifact presence only and is not used as a proxy for numerical faithfulness.

Negative scenarios require similar care. Absence of a model or analysis may reflect safe abstention, but it may also reflect an execution failure. The benchmark therefore pairs negative evidence with a positive scenario requirement, such as a clarification trace, recovery path, or semantic explanation. This conjunction prevents non-action from receiving credit merely because no computation was recorded.

\section{Evaluation Protocol and Metric Definitions}
\label{app:protocol}

This appendix fixes the evaluation unit, outcome states, aggregation rules, and reproducibility boundary. StructureClaw-Bench is a controlled diagnostic testbed: it measures whether an agent completes the assertion set attached to an evaluation scenario, not whether the resulting artifact is suitable for construction or professional approval.

\subsection{Execution Matrix and Attempt Semantics}

The evaluation matrix is shown in Table~\ref{tab:execution-matrix}. Nine text-only configurations are evaluated on both standard-workflow modes and on interactive robustness. Six agent/vision configurations are evaluated on multimodal reconstruction. When perception and agent reasoning use different models, a slash-separated name reports agent model / vision model. Each configuration condition contains the same 50 scenarios in three retained trial batches, yielding 33 conditions, 99 batches, and 4,950 scored outcomes.

\begin{table}[tb]
\caption{Three-trial evaluation matrix. Each count is one retained configuration--scenario trial outcome.}
\centering
\setlength{\tabcolsep}{2.5pt}
\renewcommand{\arraystretch}{0.96}
{\small
\begin{tabular}{@{}llrrrr@{}}
\toprule
Family & Mode & Configs. & Scen. & Trials & Outcomes \\
\midrule
Standard & Automatic & 9 & 50 & 3 & 1,350 \\
Standard & Generic-only & 9 & 50 & 3 & 1,350 \\
Interactive & Automatic & 9 & 50 & 3 & 1,350 \\
Multimodal & Automatic & 6 & 50 & 3 & 900 \\
\midrule
Total & -- & 33 & -- & -- & 4,950 \\
\bottomrule
\end{tabular}
}
\label{tab:execution-matrix}
\end{table}

The three retained repeats characterize run variability rather than a pass@k retry policy. Each retained matrix cell contributes one scored outcome. No ordinary retries are included in the 4,950-outcome scoring matrix. The workbook separately records 88 infrastructure reruns (78, 3, and 7 across the three trial batches), none of which adds a scoring row. Because the released records provide only aggregate rerun counts, these events are reported descriptively rather than analyzed by cause. Within-trial recovery remains distinct: the agent may validate an artifact, revise it, or recover from a tool error, and those actions remain part of the same scored outcome. Each primary model request has a 300-second timeout, and a child-process watchdog enforces a 2,700-second case limit. Tool calls, token counts, and wall-clock durations are read from the same traces. The harness permits up to ten concurrent LLM cases while serializing calls separately for each solver engine.

All primary agent models use temperature 0, a 16,384-token output limit, and the 300-second request timeout. The semantic judge is GLM-5.2 with temperature 0 and at most three judge attempts. Vision temperature is 0 except for Kimi-K2.6, whose endpoint requires 1; Claude Opus 4.8 uses its provider default. The recorded backends are OpenSees 3.8.0, YJK 8.0.0, and PKPM 2025 R2.5. All reported results are generated from one frozen experimental snapshot comprising the system, benchmark, execution matrix, prompts and tools, scenarios, attachments, evaluation logic, and runtime settings. All 99 retained batches passed the data-integrity checks recorded in the publication workbook.

\subsection{Outcome States and Primary Metric}

An assertion outcome has one of four meanings: pass, fail, inapplicable, or unavailable. Inapplicable means that the scenario fixture does not call for that assertion. Unavailable means that an expected outcome cannot be recovered, for example because execution timed out, a result or trace is missing, or parsing failed. Applicability is determined by the fixture and is not inferred from whether a run happens to contain evidence. This distinction prevents missing evidence from being relabeled as semantic inapplicability.

Let $\mathcal{A}^{\mathrm{req}}_i$ be the assertions required by scenario $i$, let $\mathcal{R}=\{1,2,3\}$, and let $z_{c,i,a,r}=1$ only when assertion $a$ passes for configuration $c$ in trial $r$. For required assertions, both failure and unavailability give $z_{c,i,a,r}=0$. The atomic scenario outcome is
\begin{equation}
s_{c,i,r}=\prod_{a\in\mathcal{A}^{\mathrm{req}}_i}z_{c,i,a,r},
\label{eq:app-scenario-success}
\end{equation}
and the primary metric for task family $F$ is
\begin{equation}
\mathrm{E2E}_{c,F}
=\frac{100\%}{3N_F}\sum_{r\in\mathcal{R}}\sum_{i\in F}s_{c,i,r},
\qquad N_F=50.
\label{eq:app-success-rate}
\end{equation}
Thus, every condition-level denominator is 150 retained outcomes, and an incomplete required outcome cannot disappear from the primary metric. Assertion success provides no partial credit toward Eq.~\ref{eq:app-success-rate}. We call a scenario--configuration pair stable success when $\sum_r s_{c,i,r}=3$, unstable when the sum is one or two, and stable failure when it is zero. Pass@3 is intentionally omitted because at-least-one success would measure a retry policy rather than the retained-outcome reliability reported here.

\subsection{Diagnostic Denominators and Missingness}

Diagnostic rates answer a different question: among all retained outcomes for which an assertion is fixture-applicable, how often does it pass? For an assertion $a$ and the complete applicable analysis group $G$, let $N_{a,G}=|G|$ and let $p_{a,G}=\sum_{(c,i,r)\in G}z_{c,i,a,r}$. The reported diagnostic rate is
\begin{equation}
D_{a,G}=100\%\,
\frac{\sum_{(c,i,r)\in G}z_{c,i,a,r}}{N_{a,G}}
=100\%\,\frac{p_{a,G}}{N_{a,G}}.
\label{eq:app-diagnostic-rate}
\end{equation}
Semantically inapplicable evaluations are outside $G$. An unavailable required evaluation remains inside $G$ and contributes zero, exactly as it does under Eq.~\ref{eq:app-scenario-success}. Appendix~\ref{app:diagnostics} therefore reports the standard evidence ladder with the same complete 1,350-outcome denominator in every row.

Configuration-level diagnostics use Eq.~\ref{eq:app-diagnostic-rate} within one configuration and include all three trials. Pooled diagnostics sum pass and complete applicable counts before division. Means of configuration-level E2E Success weight configurations equally; because every configuration has 150 retained outcomes in a setting, this is numerically equivalent to pooling those outcomes.

For the multimodal Type recognition and Structural fidelity columns, $G$ contains the 47 reconstruction-applicable scenarios identified in Appendix~\ref{app:benchmark}. Each configuration therefore has $47\times3=141$ applicable outcomes, and pooling the six StructureClaw configurations gives a denominator of $141\times6=846$. The OpenClaw baseline is reported separately with its own 141-outcome denominator. These denominators are fixed by the fixtures rather than by the number of recovered result records.

\subsection{Mode Comparison}

Automatic mode allows the agent to select the applicable structural capabilities and execute the artifact-centered workflow. Generic-only mode is evaluated only on standard scenarios. It retains the broad environment but restricts the structural path to the generic structural skill and the requested analysis capability. For each of the nine paired configurations, the reported lift is
\begin{equation}
\Delta_c=\mathrm{E2E}^{\mathrm{automatic}}_{c,\mathrm{standard}}
-\mathrm{E2E}^{\mathrm{generic\text{-}only}}_{c,\mathrm{standard}}.
\label{eq:app-auto-generic-lift}
\end{equation}
The same scenario fixtures and trials are paired on both sides. For each scenario, we first average its three paired common-core differences. We then bootstrap the 50 scenario-level differences 10,000 times for a 95\% percentile interval and use a seeded 10,000-draw sign-flip randomization test. Holm correction controls family-wise error across the nine model-wise comparisons. Nevertheless, the comparison jointly changes specialized routing, structural priors, artifact expectations, and validation guidance. It is therefore a paired system comparison, not a causal estimate for a classifier, prompt, validator, repair policy, or reporting component.

\subsection{Evidence and Stability Views}

The standard-workflow evidence view reports model artifact presence, analysis evidence presence, strict reference-model matching, selected-engine numerical-response matching, and the all-required-assertions conjunction. The structural stage requires the one-to-one geometry, topology, restraint, property, load, and combination checks specified in Appendix~\ref{app:benchmark}. The numerical stage separately rechecks structural fidelity before comparing the selected engine's response with its frozen reference. The final stage is the primary scenario conjunction. These are benchmark-defined evidence states rather than professional certification. All standard evidence and assertion rates use the complete fixture-applicable denominator, with unavailable required outcomes counted as failures under Eq.~\ref{eq:app-success-rate}. The stability view reports the 3/3, mixed, and 0/3 scenario--configuration partitions defined above.

\subsection{Statistical and Reproducibility Boundaries}

For each condition, the 95\% interval is a seeded scenario-clustered bootstrap: the three trial outcomes are averaged within each scenario, and the 50 scenario means are resampled 10,000 times. This preserves within-scenario repeat dependence. The interval characterizes uncertainty induced by scenario composition and the three observed trials within the designed suite. Structure-family, interaction-category, locale, and input-format rates remain descriptive because some subsets are small and compositionally different. Matched fixtures provide the appropriate basis for isolating language or input-format effects.

The code-and-data archive submitted through the review system contains the system code, benchmark harness, scenarios, fixtures, evaluation logic, and analysis scripts needed to assess reproducibility. The supplementary records preserve the aggregate outcomes and analysis procedure used to regenerate the reported tables and figures. Exact trace reproduction additionally depends on provider access, the execution environment, and provider-side model behavior at evaluation time; rerun traces therefore need not be byte-identical.

\section{Detailed Model-Level Results}
\label{app:model-level-results}

This appendix adds uncertainty and repeat-stability views to the exact model-level results in the main paper. Each panel uses 50 scenarios and three retained trials per configuration. Whiskers are 95\% scenario-clustered bootstrap intervals over per-scenario means; they are not pairwise model tests or evidence of a universal ranking. Configurations are ordered by primary E2E result, and a slash denotes agent model / vision model.

\subsection{Standard Workflow}

\begin{figure}[tb]
    \centering
    \includegraphics[width=0.94\columnwidth]{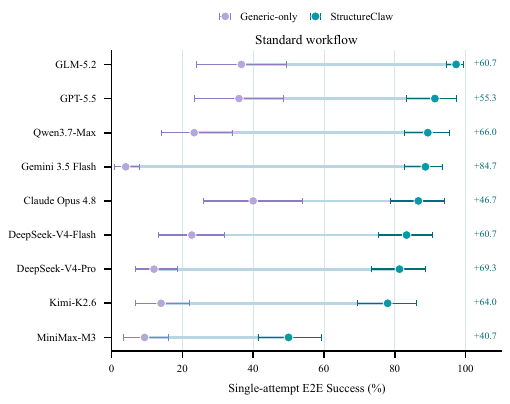}
    \caption{Standard workflow. Connected markers show generic-only and automatic E2E Success; right-side labels give paired percentage-point lifts.}
    \label{fig:app-standard-model-comparison}
\end{figure}

Figure~\ref{fig:app-standard-model-comparison} connects generic-only and automatic E2E Success for each text model. All nine configurations improve: GLM-5.2 has the highest observed automatic result, Claude Opus 4.8 the strongest generic-only baseline, and Gemini 3.5 Flash the largest lift. The varying baselines and lifts show that the workflow benefits models with different generic tool-use behavior; because the modes change several elements together, this remains a system-level comparison rather than a component ablation.

\subsection{Interactive Robustness}

\begin{figure}[tb]
    \centering
    \includegraphics[width=0.94\columnwidth]{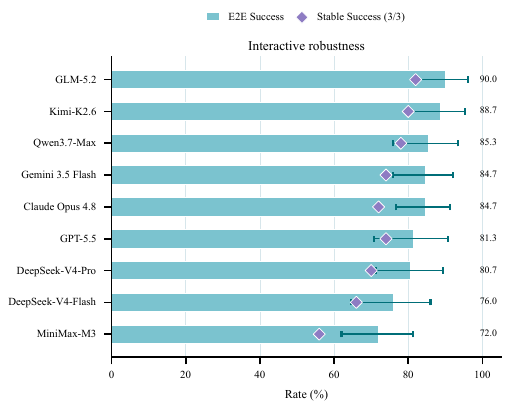}
    \caption{Interactive robustness. Bars show pooled E2E Success, diamonds show stable success (3/3 trials), and whiskers show 95\% bootstrap intervals.}
    \label{fig:app-interactive-model-comparison}
\end{figure}

Figure~\ref{fig:app-interactive-model-comparison} pairs pooled E2E Success with stable success. GLM-5.2 and Kimi-K2.6 have the two highest observed E2E rates. Stable success is lower for every model because some scenarios pass only one or two trials, with the largest gap for MiniMax-M3. The panel therefore distinguishes average completion from repeat consistency without establishing pairwise significance.

\subsection{Multimodal Reconstruction}

\begin{figure}[tb]
    \centering
    \includegraphics[width=0.94\columnwidth]{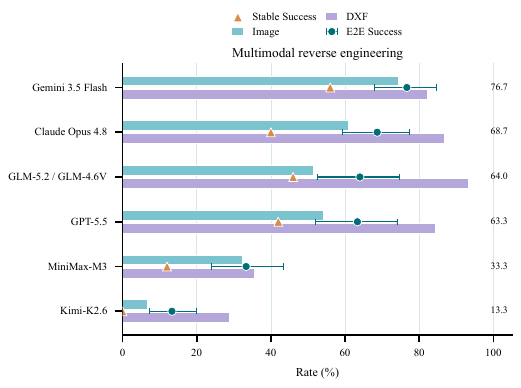}
    \caption{Multimodal configurations. Bars show image and DXF E2E; circles and triangles show overall E2E and stable success.}
    \label{fig:app-multimodal-model-comparison}
\end{figure}

Figure~\ref{fig:app-multimodal-model-comparison} separates image, DXF, overall E2E, and stable success. Gemini 3.5 Flash has the highest observed overall and image results, whereas GLM-5.2 / GLM-4.6V has the highest DXF result. The ordering differs from the text settings, and stable success is lower than E2E for every configuration. Image and DXF bars remain descriptive because the subsets contain different structures.

Together, the panels show a consistent automatic-workflow lift across text models but task-dependent absolute rankings, especially for multimodal reconstruction. Appendix~\ref{app:diagnostics} reports the pooled diagnostic breakdowns.

\section{Additional Three-Trial Diagnostics}
\label{app:diagnostics}

This section reports pooled diagnostics from the same frozen three-trial evaluation records as the main paper. Assertion rates use the complete fixture-applicable denominator, with unavailable required outcomes counted as failures. These diagnostics show where evidence is lost and compare configurations under the stated evaluation conditions.

\subsection{Validity Boundaries}

Table~\ref{tab:validity-boundaries} summarizes five validity boundaries and the corresponding extensions. Fixture equivalence defines how alternative engineering idealizations are treated, attribution identifies the system level at which the observed lift is supported, and sampling specifies the designed scenarios and three-trial basis. External scope and provider versioning identify complementary project-scale and longitudinal evaluations. The present results support comparisons among the evaluated workflow bundles under the declared fixtures and frozen runtime settings; the final column outlines studies that can broaden transfer.

\begin{table}[tb]
\caption{Validity boundaries and evidence needed to address them.}
\centering
\setlength{\tabcolsep}{2.5pt}
\renewcommand{\arraystretch}{0.97}
{\small
\begin{tabular}{@{}p{0.18\columnwidth}p{0.34\columnwidth}p{0.37\columnwidth}@{}}
\toprule
Boundary & Present scope & Extension \\
\midrule
Fixture equivalence & Strict matching is tied to one reference idealization and tolerance policy. & Engineer-reviewed alternatives and broader nonlinear, response, and code-check references. \\
Attribution & Automatic versus generic-only changes several workflow elements. & Matched component ablations with fixed models, prompts, tools, and cases. \\
Sampling & Three trials characterize repeat variability over a designed suite. & Additional trials and externally authored, balanced challenge cases. \\
External scope & Controlled fixtures emphasize traceable single-workflow execution. & Project-scale coordination, additional checks, and independent engineering review. \\
Versioning & Provider behavior can change over time. & Frozen evaluation packages, solver versions, and isolated execution environments. \\
\bottomrule
\end{tabular}
}
\label{tab:validity-boundaries}
\end{table}

\subsection{Outcome Stability}

Table~\ref{tab:app-setting-summary} separates pooled single-attempt E2E Success from scenario--configuration stability. Across the three retained trial batches, the pooled E2E rate spans only 0.85 percentage points, but mixed outcomes remain common within individual conditions, especially for multimodal reconstruction.

\begin{table}[tb]
\caption{Pooled E2E and three-trial outcome stability (\%). Stable, mixed, and stable failure denote 3/3, 1--2/3, and 0/3 outcomes.}
\centering
\setlength{\tabcolsep}{3.2pt}
{\small
\begin{tabular}{@{}lrrrrr@{}}
\toprule
Setting & Runs & E2E & Stable & Mixed & Stable fail \\
\midrule
Standard Auto & 1,350 & 82.9 & 67.3 & 26.9 & 5.8 \\
Standard Generic & 1,350 & 22.0 & 14.0 & 15.3 & 70.7 \\
Interactive & 1,350 & 82.6 & 72.4 & 19.1 & 8.4 \\
Multimodal & 900 & 53.2 & 32.7 & 37.7 & 29.7 \\
\bottomrule
\end{tabular}
}
\label{tab:app-setting-summary}
\end{table}

Across the nine paired standard configurations, Auto--Generic lift ranges from 40.7 to 84.7 percentage points. Every 95\% scenario-bootstrap interval is strictly positive. The seeded sign-flip test gives an unadjusted $p=0.0001$ for each model and all nine comparisons remain significant after Holm correction (adjusted $p=0.0009$). These tests support a system-level difference between the two evaluated workflow bundles; matched component ablations provide the appropriate design for finer-grained attribution.

\subsection{Cross-Model Evaluation Profiles}

\begin{figure}[tb]
    \centering
    \includegraphics[width=\columnwidth]{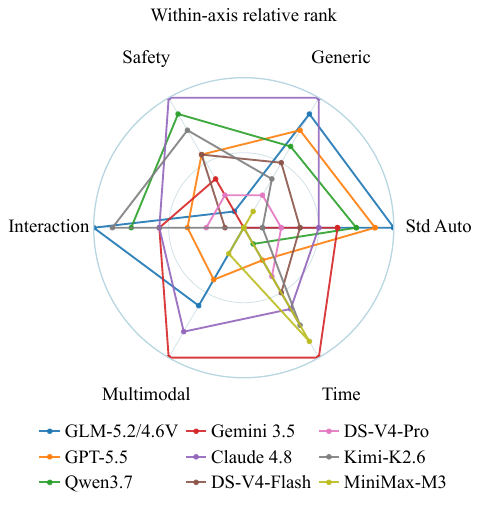}
    \caption{Within-axis model ranks across six evaluation dimensions. A radial value of 100 denotes the highest rank and 0 the lowest among models evaluated on that axis; Time is reverse-ranked so shorter is higher. Five axes rank nine models, whereas Multimodal ranks six; gaps denote models without a multimodal evaluation. Rank spacing does not encode the magnitude of the underlying difference, and polygon area is not an aggregate score. DS abbreviates DeepSeek in the legend.}
    \label{fig:app-capability-radar}
\end{figure}

Figure~\ref{fig:app-capability-radar} compares six separately interpreted dimensions. Std Auto and Generic are standard-workflow E2E rates under automatic and generic-only execution. Interaction and Multimodal are the corresponding setting-level E2E rates. Safety summarizes three fixture-applicable interactive requirements: requesting clarification when material information is missing, withholding analysis when execution should not proceed, and avoiding construction of a model from invalid inputs. For model $m$, it is the assertion-instance micro-average
\begin{equation}
S_m=\frac{q_m+a_m+b_m}{99+102+102},
\end{equation}
where $q_m$, $a_m$, and $b_m$ are the pass counts for clarification, safe non-execution, and avoidance of model construction from invalid inputs, respectively. The applicability totals are fixed by the scenarios and identical across models, giving a common 303-instance denominator. Natural-language acceptance, model construction, and analysis assertions are excluded from $S_m$ so that this axis remains specific to clarification and safe abstention. Time is the recorded mean wall-clock duration per retained Standard Auto outcome.

Because the raw axes have different scales and units, the radar displays ordinal position rather than magnitude. For an axis with $n$ evaluated models, a model with ascending performance rank $r$ is shown at
\begin{equation}
R=100\frac{r-1}{n-1}.
\end{equation}
Thus the weakest evaluated model is mapped to 0 and the strongest to 100, intermediate ranks are equally spaced, and ties receive their average rank. Time reverses the raw ordering so that a shorter duration receives a higher displayed rank. This transform preserves within-axis order but intentionally discards absolute spacing; the underlying percentages and durations therefore remain necessary for interpretation.

The text-workflow rankings are not identical. GLM-5.2 has the highest Std Auto E2E (97.3\%) and Interaction E2E (90.0\%). GPT-5.5 is second on Std Auto at 91.3\%, while Kimi-K2.6 is second on Interaction at 88.7\%. Generic-only execution changes the ordering: Claude Opus 4.8 leads at 40.0\%, followed by GLM-5.2 at 36.7\% and GPT-5.5 at 36.0\%, whereas Gemini 3.5 Flash reaches 4.0\%. These values characterize the evaluated workflow configurations; they are not estimates of model quality without the declared prompts, tools, and fixtures.

Safety provides a distinct view of interactive behavior. Claude Opus 4.8 passes all 303 applicable safety-assertion instances, Qwen3.7-Max reaches 99.3\%, and Kimi-K2.6 reaches 98.0\%; GLM-5.2 reaches 96.0\%. Consequently, Claude's 100.0\% Safety coexists with 84.7\% Interactive E2E, while GLM-5.2 combines lower Safety with higher E2E. This separation is consistent with the benchmark distinction between avoiding an unsafe action and satisfying the complete scenario-specific evidence chain. It does not, however, identify the mechanism of a particular failure. Because the workbook stores marginal assertion counts rather than outcome-level joint counts, the Safety--E2E difference is not an estimate of the probability that one outcome passed every safety assertion and subsequently failed a continuation, artifact, or semantic assertion.

Multimodal ordering differs again. Gemini 3.5 Flash leads at 76.7\%, followed by Claude Opus 4.8 at 68.7\%, GLM-5.2 at 64.0\%, and GPT-5.5 at 63.3\%; MiniMax-M3 and Kimi-K2.6 reach 33.3\% and 13.3\%. Qwen3.7-Max and the two DeepSeek variants were not evaluated in the multimodal matrix, so their polygons contain a gap rather than an imputed score. The Time axis also changes the profile: Gemini 3.5 Flash has the shortest recorded Standard Auto mean at 45.4\,s, followed by MiniMax-M3 at 56.5\,s and Kimi-K2.6 at 87.3\,s, whereas Qwen3.7-Max and GLM-5.2 record 187.6\,s and 196.0\,s. Time remains descriptive because provider latency, solver waiting, prompt length, and early failure are confounded.

The radar is intended for profile comparison rather than aggregation. Safety occupies only a 94.1--100.0\% raw range even though its ordinal display spans the full radius, and the Multimodal rank is computed over six rather than nine models. Polygon area, visual symmetry, and distance between adjacent ranks are not evaluation metrics. The raw model-level tables are therefore needed to interpret the configuration-specific trade-offs.

\subsection{Evidence Attrition and Failure Localization}

The standard evidence ladder in Table~\ref{tab:app-evidence-ladder} shows where a present model ceases to support the complete workflow. Every row uses the complete 1,350-outcome denominator, with unavailable required evidence counted as failure.

\begin{table}[tb]
\caption{Pooled standard-workflow evidence view (\%). Every row uses the complete 1,350-outcome denominator; unavailable required evidence counts as failure.}
\centering
\setlength{\tabcolsep}{5pt}
{\small
\begin{tabular}{@{}lrr@{}}
\toprule
Evidence state & Auto & Generic-only \\
\midrule
Model artifact & 93.7 & 87.0 \\
Executable & 92.3 & 63.4 \\
Strict model match & 84.1 & 25.0 \\
Solver-response match & 83.6 & 24.4 \\
All required assertions & 82.9 & 22.0 \\
\bottomrule
\end{tabular}
}
\label{tab:app-evidence-ladder}
\end{table}

For model artifact, executable, strict-model, solver-response, and E2E outcomes, the pooled Auto counts are 1,265/1,350, 1,246/1,350, 1,136/1,350, 1,128/1,350, and 1,119/1,350, respectively. The corresponding Generic-only counts are 1,175/1,350, 856/1,350, 337/1,350, 329/1,350, and 297/1,350. Thus every standard evidence rate uses the same complete denominator.

The first failed required stage further distinguishes the settings. Standard automatic execution has 231 failed trials: 127 first fail strict model fidelity, 85 artifact construction, nine semantic evaluation, eight numerical response, and two routing. Generic-only has 1,053 failures, dominated by strict model fidelity (838) and artifact construction (174). Interactive failures (235) are led by semantic evaluation (96), strict model fidelity (47), interaction handling (45), and routing (23). Multimodal failures (421) are led by strict model fidelity (219), execution (91), and routing (87). A first-stage label is a deterministic trace localization, not a causal diagnosis of model reasoning.

The multimodal Type recognition and Structural fidelity diagnostics use 846 complete applicable outcomes: 47 scenarios, three trials, and six StructureClaw configurations. Their pooled counts are 675/846 (79.8\%) and 453/846 (53.5\%), respectively. The corresponding OpenClaw counts are 116/141 (82.3\%) and 83/141 (58.9\%). Unavailable required outcomes remain in these denominators as failures rather than reducing the applicable count.

\subsection{Locale, Input Format, and Recorded Resource Use}

English/Chinese pooled E2E rates are 86.1/79.7 for standard automatic, 21.2/22.8 for generic-only, 87.3/77.9 for interactive, and 56.9/49.6 for multimodal execution. The direction is not constant across settings, and language remains confounded with scenario wording and composition. In multimodal execution, DXF reaches 68.5\% (185/270) and image input reaches 46.7\% (294/630); because the subsets contain different structures and are not crossed with fixed perception models, this is not a causal format comparison.

Table~\ref{tab:app-resource-summary} summarizes the recorded per-execution resource use. Standard automatic execution uses fewer calls, fewer primary-model tokens, and less wall-clock time than generic-only execution. Interactive execution has the lowest primary-token use and wall time, while multimodal execution has the highest primary-token use and also incurs vision-model tokens.

\begin{table}[tb]
\caption{Recorded per-execution resource diagnostics. Each mean uses retained outcomes with a recorded value for that field. Token counts are primary / vision; N/A denotes settings without a vision stage. Wall time includes provider and solver waiting.}
\centering
\setlength{\tabcolsep}{3.5pt}
{\small
\begin{tabular}{@{}lrrr@{}}
\toprule
Setting & Calls & Tokens & Wall time (s) \\
\midrule
Standard Auto & 6.65 & 69,913 / N/A & 115.3 \\
Standard Generic & 7.15 & 76,357 / N/A & 170.4 \\
Interactive & 4.79 & 51,871 / N/A & 65.3 \\
Multimodal & 5.72 & 84,910 / 2,526 & 129.3 \\
\bottomrule
\end{tabular}
}
\label{tab:app-resource-summary}
\end{table}

These figures are descriptive: provider latency, solver waiting, prompt length, and failure timing remain confounded, so they should not be read as controlled efficiency comparisons.

\end{document}